\documentclass[twocolumn,twoside,slac_two]{revtex4}
\usepackage{graphicx}
\usepackage{fancyhdr}
\usepackage{graphics}
\usepackage{epstopdf}
\usepackage{textpos}
\begin{document}

\title{\centering The Mu2e Experiment at Fermilab}


\author{
\centering
\begin{center}
Robert K. Kutschke
\end{center}}
\affiliation{\centering Fermi National Accelerator Laboratory, Batavia Illinois, 60542, USA}
\begin{abstract}
The Mu2e collaboration has proposed an experiment
to search for the coherent, neutrino-less conversion of a muon into an electron in the 
Coulomb field of a nucleus
with an expected sensitivity of $R_{\mu e}<6.0\times 10^{-17}$, at the 90\% confidence level.
Mu2e has received CD-0 approval from the US Department of Energy.  If all resources are made available as required, the  experiment could begin taking data as early as 2018.
\end{abstract}

\maketitle
\thispagestyle{fancy}


\section{INTRODUCTION}

Within the Standard Model, muons decay in a way that almost perfectly conserves lepton family number; the standard model branching fractions to final states that violate lepton family number are much too small to be observed with present technology.  Therefore any observation of a muon decay that violates lepton family number is direct evidence for physics beyond the Standard Model.   One of the most  striking of these decay modes cannot occur for free muons but may  occur when negative muons are bound to an atomic nucleus, forming a muonic atom:  coherent,  neutrino-less muon to electron conversion in the Coulomb field of a  nucleus.  In the process the initial state is an muonic atom
and  the final state is a mono-energetic electron plus an unobserved,  recoiling, intact nucleus; there are no neutrinos in the final state. 

The Mu2e collaboration has proposed an experiment to search for muon to electron conversion; this experiment is to be mounted at Fermilab and, if no events are seen in the signal window,  the experiment is designed to set an upper limit of $R_{\mu e} \le 6\times 10^{-17}$ at the 90\% confidence level.   The
ratio f $R_{\mu e}$ is defined by,
\begin{eqnarray}
R_{\mu e} &=& \frac
{\Gamma(\mu^{-}\;  N(A,Z) \to e^{-}\; N(A,Z)}
{\Gamma(\mu^{-}\; N(A,Z)\to \nu_{\mu}\; N(A,Z-1))},
\end{eqnarray}
where $N(A,Z)$ denotes a nucleus with mass number $A$ and atomic number $Z$.  The numerator is the rate for the conversion process and the denominator is the rate for ordinary muon capture on
the same nucleus.

In the standard model, muon to electron conversion can proceed via a penguin diagram that contains a 
$W$ and an oscillating neutrino in the loop; the $W$ exchanges a photon with
the nucleus.   Estimates of the standard model rate for this process predict that $R_{\mu e}$ is of order $10^{-57}$.  Most new physics scenarios predict much
larger values of $R_{\mu e}$; in particular,
 scenarios that predict that SUSY is within the reach of the LHC also predict
 $R_{\mu e}\approx 10^{-15}$, a rate for which Mu2e would observe about 40 events on
 a background of fewer than $0.2$ events.

   The process of muon to electron conversion is just one example in the broader field of Charged
 Lepton Flavor Violation (CLFV). 
 An excellent review of CLFV and the flavor physics of leptons can be found in reference~\cite{review}.
Two classes of diagrams can contribute  to conversion.  The first class includes magnetic moment 
loop diagrams, with a photon exchanged between the loop and the nucleus; these diagrams can
proceed with many different sorts of particles in the loop, including, but not limited to,
SUSY particles, heavy neutrinos and a second Higgs Doublet.    This class of diagrams also produces
non-zero rates for the process $\mu\to e\gamma$.
The second class of diagrams includes both contact terms, which parameterize compositeness, and
the exchange of a new heavy particle, such as a Leptoquark or a $Z^{\prime}$.  This class of diagrams does not give rise to the process $\mu\to e\gamma$.
Through these processes, Mu2e has sensitivity to mass scales up to
10,000 TeV, far beyond the scales that will be accessible to direct observation at
the LHC.  Mu2e also has access to physics that cannot be probed by $\mu\to e\gamma$.

\section{THE EXPERIMENTAL TECHNIQUE}
The Mu2e apparatus is described in detail in the Mu2e proposal~\cite{proposal}.
The basic idea behind Mu2e is motivated by the MECO~\cite{MECO} experiment,
which, in turn, was motivated by the MELC experiment.
A beam of low momentum negative muons is stopped on a set of thin  aluminium target foils and
the muons drop to the K-shell, forming a muonic atom.    Mu2e will measure the rates
of the characteristic X-rays emitted in this cascade, which is  a part of establishing 
the denominator in $R_{\mu e}$.
The two major decay modes of muonic Al are muon capture on the nucleus, which occurs 
approximately 60\% of the time, and muon decay in orbit (DIO), which occurs approximately 
40\% of the time.
Muon nuclear capture produces, protons, neutrons and photons; these produce hits
in the detector with enough rate to complicate pattern recognition but they 
produce fake signal electrons only via secondary processes.

\begin{figure}
\includegraphics[width=50mm]{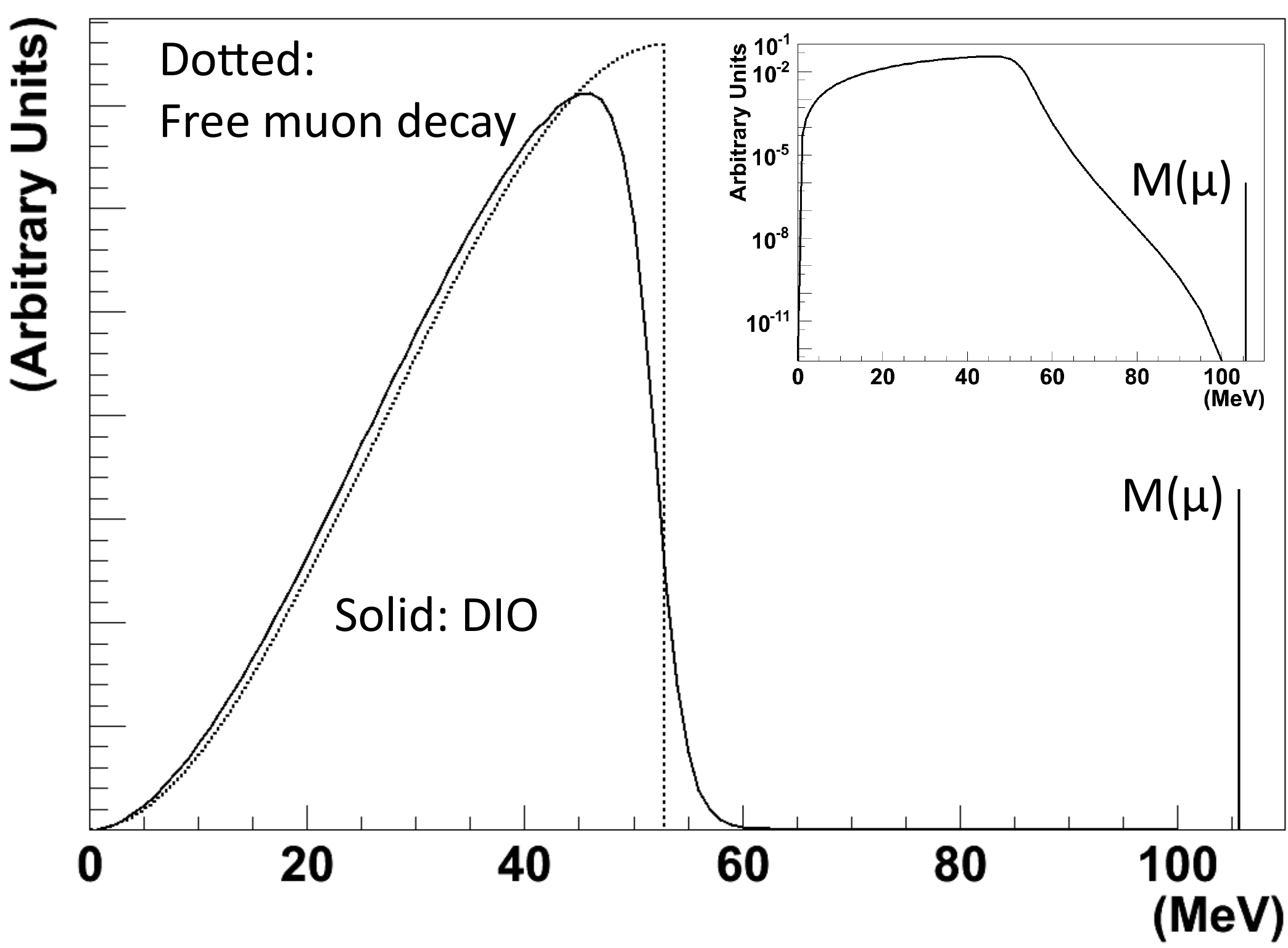}
\caption{Electron energy spectrum from muon decay in orbit.
In the main figure the dotted curve shows the Michel spectrum from the decay
of a free muon; this has a hard endpoint at half of the muon mass. The solid curve 
shows the spectrum from the decay in orbit of muonic aluminium, as calculated
by reference~\cite{DIOWatanabe};
the shape of this spectrum is  a distorted Michel spectrum with a long tail to high energies.  
The inset shows this DIO spectrum on a log scale; the spectrum extends all of the way out to 
the endpoint energy,  about 0.5 MeV less than the muon mass.  
For reference, the muon mass is drawn on both figures.
}
\label{fig:dio}
\end{figure}

DIO produces electrons with a 
continuous energy spectrum, as shown in Figure~\ref{fig:dio};
the shape is a distorted Michel spectrum with a long tail, due to  
radiative corrections in which the outgoing electron coherently exchanges a photon with
the nucleus.
In one extreme  configuration, both neutrinos are at rest and the electron recoils against
the intact Al nucleus.  This is the configuration in which the electron has the maximum energy in 
the lab frame, about 105~MeV for muonic Al.    This energy is equal to the muon mass, less a small
correction for the K-shell binding energy and an even smaller correction for nuclear recoil.

The $\mu$ to $e$ conversion process produces a mono-energetic electron with an energy equal to
that of the endpoint of the continuous spectrum from DIO.  An irreducible background comes
from electrons
in the high energy tail of the DIO spectrum that are mis-measured with a momentum in the signal
region; Mu2e addresses this background by designing a tracking system that measures the momentum to one part in 1000.
In summary, the Mu2e experimental technique is to carefully
measure the energy spectrum from electrons emitted from the target foils and to search for
an excess at the endpoint.

\begin{figure*}[t]
\centering
\includegraphics[width=135mm]{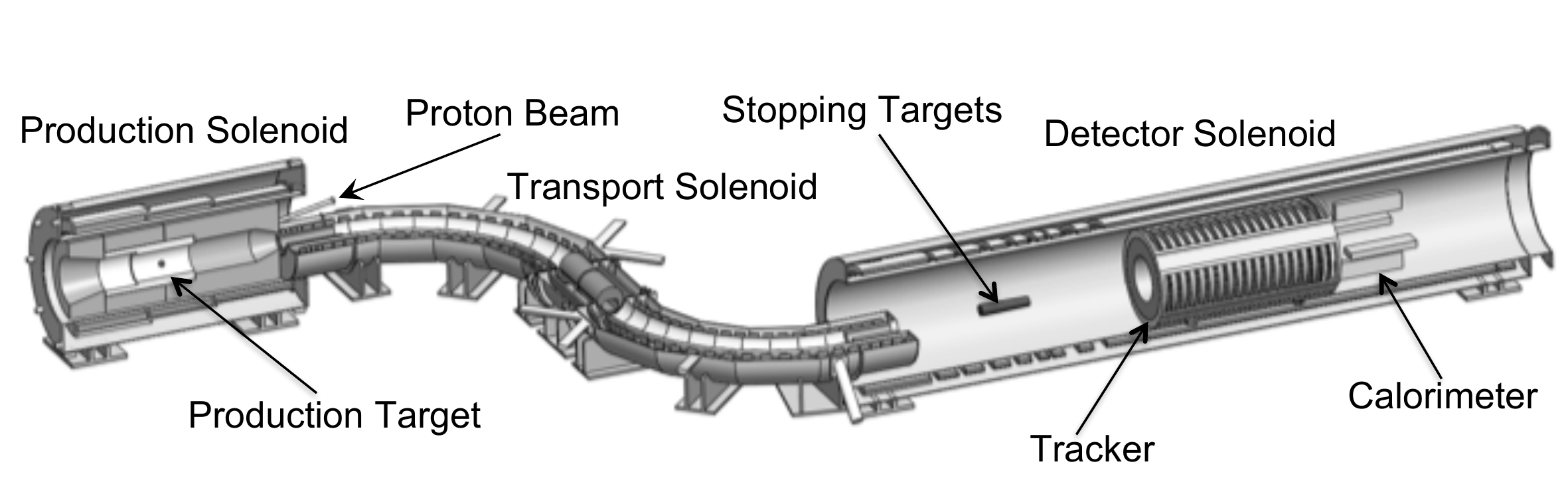}
\caption{Diagram of the Mu2e muon beam-line and detector.  The proton beam enters from the right at the left side of the figure.  A back-scattered muon beam is captured by the Production Solenoid and transported through the S-bend Transport Solenoid to the stopping targets.  Conversion electrons, produced in the stopping target are captured by the magnetic field in the Detector Solenoid and transported through the Tracker, which makes a precision measurement of the momentum.  The conversion electrons then strike the Electromagnetic Calorimeter, which provides independent, confirming measurements.  Not shown in this figure are the cosmic ray veto system and
the muon stopping monitor.}
\label{fig:mu2e}
\end{figure*}

The muon beam used by Mu2e is produced using 8 GeV protons from the Fermilab accelerator 
complex.  In order to minimize construction costs, Mu2e will reuse many parts of the accelerator 
complex following the completion of  Tevatron Run II.
A bunch of protons with a full width of about $100$~ns is steered onto a pencil shaped
Au target located in the middle of a high field, graded-field 
solenoid, the Production Solenoid (PS), which is shown in Figure~\ref{fig:mu2e}.
In the production target, p-Au interactions produce pions
that are captured into helical trajectories in the field of the solenoid; these pions decay into muons 
that are  also captured  by the field of the solenoid.  Mu2e collects the backscattered muon beam 
and transports it into the Transport Solenoid (TS), also shown in Figure~\ref{fig:mu2e}.
The PS has a graded magnetic field,
with a field of 5~T at the proton-downstream end, falling to about 2.5~T at the proton-upstream end. 
This forms a magnetic mirror that reflects a portion of the forward going pions and muons, 
increasing the yield of captured muons.

The TS has an S-bend that induces a dipole term in the magnetic field; this allows, by appropriate placement of absorbers and collimators, the sign selection of the muon beam.  A thin window in the path
of the negatively charged beam stops most of the anti-proton contamination whilst transmitting almost
all of the muons.
The TS transmits the $\mu^{-}$ beam into the Detector Solenoid (DS) where it encounters the 17 thin
Al foils that comprise the stopping target.  About 50\% of the muons range out in one of the foils and are
captured to form muonic Aluminium.
Downstream of the target is a tracking system and
downstream of that is an electromagnetic calorimeter (ECal).  In both of these devices, the inner region,
out to a radius of about 38~cm is empty.  This allows those
muons that do not stop in the stopping target to pass through the detector to the muon beam dump.
This also permits the low $p_T$ subset of the background particles to pass through the detector
to the muon beam dump.

The DS magnetic field is also graded 
to form a magnetic mirror that reflects some backwards going electrons towards the tracker.
In the volume occupied by the tracker and ECal, the DS magnetic field is highly uniform at 1.0~T.  
When a conversion or DIO electron is emitted from the stopping target, it travels in a helical trajectory
and, if it has sufficient transverse momentum, $p_{T}$, its trajectory will be measured by the tracker.
Only those electrons with $p_{T}>55$~MeV$/c$ will reach the tracker and only those with
$p_{T}>90$~MeV$/c$ will intersect enough of the tracker to form a reconstructible track.
Because almost all tracks from DIO have $p_{T}<m_{\mu}/2$ they will never reach the tracker.
This is the key to making a measurement of $R_{\mu e}$ with a sensitivity of $O(10^{-17})$: 
the apparatus is only sensitive to the tail of the DIO energy distribution. 

The $\mu^{-}$ beam that reaches the stopping target is contaminated by many $e^{-}$
and some $\pi^{-}$, both of which can produce false signals when they interact with the 
stopping targets.   These backgrounds occur promptly.  To defeat them,  the experiment
exploits the lifetime of muonic Al, about 864~ns: Mu2e waits for 770~ns following the arrival
of the proton bunch at the production target and then begins counting electrons that
are emitted from the foils of the stopping target.   By this time, all of the beam from the production
target has passed through the stopping target and the prompt backgrounds have died away.
After a total of 1694~ns the cycle is repeated.

It is also critical that few protons arrive at the production
target between the bunches.
If protons arrive out of time, they can produce $e^{-}$ and $\pi^{-}$ that arrive at the stopping
target within the live gate.  To reduce this background Mu2e requires an
extinction of $10^{-10}$; that is,
for every proton that arrives at the production
target within the bunch, there should be no more 
than $10^{-10}$ protons between bunches.  

Several processes can produce a true electron that can be (mis-)measured to have
an energy in the signal region.   The dominant sources of such false signal
electrons are expected to be
mis-measured DIO electrons  ($0.009\pm0.006$ events),
radiative $\pi^{-}$ capture on the target foils ($0.04\pm0.02^{\ddag}$ events),
$\mu$ decay in flight ( $0.034\pm0.017^{\ddag}$ events), and
cosmic ray induced   ($0.025\pm0.025$ events).  These and
seven other processes add to 
total estimated background of $0.17\pm0.007$ events.  These numbers are quoted for
$3.6\times 10^{20}$ protons on target.
The processes marked $^{\ddag}$ scale with extinction and are reported for
an extinction of $10^{-10}$.

 The critical path for building the Mu2e apparatus is the design and construction of the solenoid system.
 If all resources are made available as required, the solenoids could be
 installed by 2018.  At CD-0, the collaboration estimated  a total project cost on the order of
 M\$200.
 
\section{SUMMARY AND CONCLUSIONS}
The goal of the Mu2e experiment is to observe $\mu$ to e conversion or to set an
upper limit of $R_{\mu e}< 6\times 10^{-17}$ at the 90\% CL, which will require
$3.6\times10^{20}$ protons on target.
This sensitivity is 10,000 times better than the previous best limit~\cite{SINDRUM};
mass scales up
to O(10,000~TeV) are within reach.  For $R_{\mu e}=10^{-15}$ Mu2e would measure about
40~events on a background of less than 0.2 events.  The experiment has received
CD-0 from the US Department of Energy and will soon be reviewed for CD-1.
Visit the Mu2e home page~\cite{home} to keep up to date with the experiment.

\bigskip 
\begin{acknowledgments}
The author would like to thank the Fermilab management and staff for their strong support of the 
Mu2e experiment. This work is supported in part by the U.S. Department of 
Energy, the U.S. National Science Foundation, INFN in Italy, and the 
U.S.-Japan Agreement in High Energy Physics.
\end{acknowledgments}

\bigskip 
\bibliography{basename of .bib file}

\begin{thebibliography}{9}   
%
%
%

\bibitem{review}
M.~Raidal {\it et al.}, \emph{Eur. Phys. J. C57}:13-182, 2008 (arXiv:0801.1826).

\bibitem{proposal} Mu2e-doc-388.   All Mu2e documents can be retrieved
from the Mu2e Document Database, \\
http://mu2e-docdb.fnal.gov/\\
cgi-bin/DocumentDatabase

\bibitem{MECO} 
The MECO home page:\\
http://www.bnl.gov/rsvp/MECO.htm

\bibitem{DIOWatanabe}
R.~Watanabe {\it et al.}, \emph{Atomic Data and Nuclear Data Tables 54}:165,   1993.

\bibitem{SINDRUM}
W. Bertl {\it et al.}  (SINDRUM II Collab.) \emph{Eur. Phys. J. C47}:337, 2006.

\bibitem{home} The Mu2e home page:
http://mu2e.fnal.gov

\end{thebibliography}

\end{document}